\documentclass[conference]{IEEEtran}
\IEEEoverridecommandlockouts

\usepackage{booktabs} 

\usepackage{url}
\usepackage{tcolorbox}

\usepackage{amsmath}
\usepackage{float}
\usepackage{mathrsfs}

\usepackage{booktabs}
\usepackage{graphicx} 
\usepackage{multirow}

\usepackage{cite}
\usepackage{amsmath,amssymb,amsfonts}
\usepackage{algorithmic}
\usepackage{graphicx}
\usepackage{textcomp}
\usepackage{xcolor}

\def\BibTeX{{\rm B\kern-.05em{\sc i\kern-.025em b}\kern-.08em
    T\kern-.1667em\lower.7ex\hbox{E}\kern-.125emX}}
\begin{document}

\title{A test-free semantic mistakes localization framework in Neural Code Translation\\
}
\author{
\IEEEauthorblockN{1\textsuperscript{st} Lei Chen}
\IEEEauthorblockA{\textit{School of Computer Science and Technology} \\
\textit{Tianjin University}\\
Tianjin, China \\
2022244117@tju.edu.cn}
\and
\IEEEauthorblockN{2\textsuperscript{nd} Sai Zhang}
\IEEEauthorblockA{\textit{School of Computer Science and Technology} \\
\textit{Tianjin University}\\
Tianjin, China \\
zhang\_sai@tju.edu.cn}
\and
\IEEEauthorblockN{3\textsuperscript{rd} Fangzhou Xu}
\IEEEauthorblockA{\textit{School of Computer Science and Technology} \\
\textit{Tianjin University}\\
Tianjin, China \\
xu\_fangzhou@tju.edu.cn}
\and
\IEEEauthorblockN{4\textsuperscript{th} Zhenchang Xing}
\IEEEauthorblockA{\textit{CSIRO's Data61} \\
\textit{CSIRO}\\
Australia \\
zhenchang.xing@data61.csiro.au}
\and
\IEEEauthorblockN{5\textsuperscript{th} Liang Wan}
\IEEEauthorblockA{\textit{School of Computer Science and Technology} \\
\textit{Tianjin University}\\
Tianjin, China \\
lwan@tju.edu.cn}
\and
\IEEEauthorblockN{6\textsuperscript{th} Xiaowang Zhang\textsuperscript{*}}
\IEEEauthorblockA{\textit{School of Computer Science and Technology} \\
\textit{Tianjin University}\\
Tianjin, China \\
xiaowangzhang@tju.edu.cn}
\and
\IEEEauthorblockN{7\textsuperscript{th} Zhiyong Feng}
\IEEEauthorblockA{\textit{School of Computer Science and Technology} \\
\textit{Tianjin University}\\
Tianjin, China \\
zyfeng@tju.edu.cn}
}


\maketitle

\begin{abstract}
In the task of code translation, neural network-based models have been shown to frequently produce semantically erroneous code that deviates from the original logic of the source code. This issue persists even with advanced large models. Although a recent approach proposed using test cases to identify these semantic errors, it relies heavily on the quality of the test cases and is not applicable to code snippets without test cases in real-world scenarios. Therefore, We present EISP, a static analysis framework based on the Large Language Model (LLM).First, the framework generates a semantic mapping between source code and translated code. Next, each sub-code fragment is identified by recursively traversing the abstract syntax tree of the source code, and its corresponding translated code fragment is found through the semantic mapping. Finally, EISP connects each pair of sub-code fragments with fine-grained knowledge hints through an AI chain to assist LLMs in discovering semantic mistakes in the translated code. In our benchmark evaluation, the EISP framework, based on GPT-4o mini, achieved an accuracy of 82.3\%, representing a 20.3\% improvement over baseline methods using the same base model, and a 7.4\% improvement compared to dynamic analysis methods that require test cases and manual intervention. To our knowledge, EISP is the first tool to locate semantic errors in translated code without test cases or compilable code. This innovative tool provides the software engineering community with a new way to deal with code fragments without test cases.
\end{abstract}


\section{Introduction}




Code translation is the process of converting a program from one programming language to another, with the goal of language conversion while maintaining the original functionality. Code translation technology plays a central role in modern software development, not only significantly reducing the cost of migrating legacy software from proprietary languages to more general-purpose programming languages~\cite{gholami2017challenges,bergmayr2013migrating}, but also supporting the modernization of enterprise applications~\cite{ kalia2021mono2micro,perez2021software}.


Due to the importance of code translation in modern development, a variety of techniques have been applied to this area, with neural network-based models being particularly prominent ~\cite{brown2020language,lample2018phrase,roziere2020unsupervised}. In particular, large-scale language models (LLMs), such as ChatGPT, were originally designed for natural language generation tasks, but they also perform well in code translation tasks. Moreover, code translated by neural network models is more natural and consistent with human programming habits than code generated by traditional compilers. In particular, using the latest large language models, we can enhance the readability of the code by using natural language hints to make it take into account comments or variable name naming conventions, for example.

Although neural network models have achieved significant success in code translation tasks, recent research has found that ~\cite{pan2024lost,wang2023transmap}, even with advanced large language models such as ChatGPT, these models still face challenges in automated code translation, especially when dealing with complex code. This is because code translation is a complex task that involves a deep understanding of the syntax and semantics of the code. Neural network models need to generate syntactically correct code and ensure that the original functionality is preserved during translation. However, existing models that deal with this task tend to introduce subtle errors that may cause the translated code to be semantically inconsistent with the original code. These errors are widespread and include syntax errors, incorrect handling of loops and conditional statements, data type mismatches, and mismatches between the source code and the API behavior of the translated code~\cite{pan2024lost,jain2022jigsaw,roziere2020unsupervised}. These subtle errors can be categorized into two main groups: syntactic errors and semantic errors. Syntactic errors violate the syntactic rules of the target language and can usually be recognized more easily by a syntax detector. In contrast, semantic errors are more insidious and may result in code that fails to execute without violating the target language syntax, or that executes but whose output does not match the original code. In addition, for many of these semantic errors, even running a test case does not immediately reveal the location of the fix, and such errors are also known as hidden errors ~\cite{wang2023transmap}.

Based on this, Wang et al.~\cite{wang2023transmap} developed a dynamic analysis framework that automatically locates semantic errors in code translation by executing test samples of source code and translated code, and gradually comparing the output values of each line of code. The approach requires source code, translated code with errors, and corresponding test code that needs to be able to trigger the recognized errors. The goal is to pinpoint the location of errors in the translated code, allowing the programmer to focus on fixing specific lines of code without having to analyze the entire program. However, the study by Wang et al. has some limitations. The framework is highly dependent on the quality of test samples and is unable to deal with code fragments that lack valid test samples, which somewhat limits the applicability of the approach and its ability to handle data types. In addition, recent research ~\cite{yan2023codetransocean} has shown that in the field of code translation, the approach of relying on test cases for code execution is not only costly, but may also pose a security risk. For code that is imported incompletely, even if test cases exist, the code needs to be supplemented or modified.

To automatically locate semantic errors in the translation of neural network code in the absence of test cases, we propose EISP, a static analysis framework based on the Large Language Model (LLM), which utilizes knowledge of fine-grained functional descriptions extracted from API documentation and combines it with the analytical capabilities of the LLM to decompose the source code and translated code in tandem, and then analyze each pair of sub-codes statically to thereby locate semantic errors in the translated code. Our approach consists of two parts: the offline part involves extracting knowledge from API documentation to build an API knowledge base, while the online part first generates a semantic mapping of the input source and translation code by prompting LLM with a prompt, which is then stored in a database. Then the system recursively traverses the sub-code fragments of the source code through the Abstract Syntax Tree (AST) and identifies the corresponding translated code fragments by using the semantic mappings in the database, realizing the synchronous decomposition of the source code and the translated code. Finally, each code fragment and its corresponding translated code fragment are linked to fine-grained knowledge hints through an AI chain, which in turn assists LLM to statically analyze semantic errors in the translated code.



To evaluate the effectiveness of the EISP framework, we adopted the benchmark dataset from previous work~\cite{wang2023transmap} and extended it by following the original data collection procedure to gather additional data, merging it with the existing benchmark. This resulted in a new benchmark dataset containing 786 identified and verified semantic errors. Additionally, in our experiments, we excluded the test code from each sample, using only the source code and the translated code as inputs to the EISP framework. The experimental results showed that the EISP framework, based on GPT-4o mini, successfully identified 82.3\% of the semantic errors, achieving a 20.3\% improvement in success rate compared to the baseline methods using the same base model.To further evaluate the performance of our approach with different LLMs, we replaced GPT-4o mini in EISP with GPT-3.5-turbo and llama-3-70b-Instruct-bnb-4bit, respectively, and conducted experiments on the same benchmark dataset. The results indicated that EISP, using different base models, consistently outperformed the corresponding baseline methods. Finally, we compared EISP with the dynamic analysis method TransMap. We found that EISP achieved a 7.4\% higher success rate in identifying overall semantic errors than TransMap, and outperformed TransMap without human intervention by 31\%. Furthermore, compared to TransMap, EISP does not require test cases and manual intervention, demonstrating greater generalizability and automation capabilities.

Key contributions of this work include:
\begin{itemize}
\item We are the first work to localize semantic errors in code translation for code without test cases.
\item We present a new method for locating semantic errors in code translation that requires only API documentation, does not require training of the model, and is adaptable to low-resource languages.
\item The experimental results demonstrate that our approach outperforms existing baseline methods, including dynamic analysis methods that rely on test cases and manual intervention.
\end{itemize}

\section{Motivation And Related Work}

\begin{figure*}[h]
  \centering
  \includegraphics[width=\linewidth]{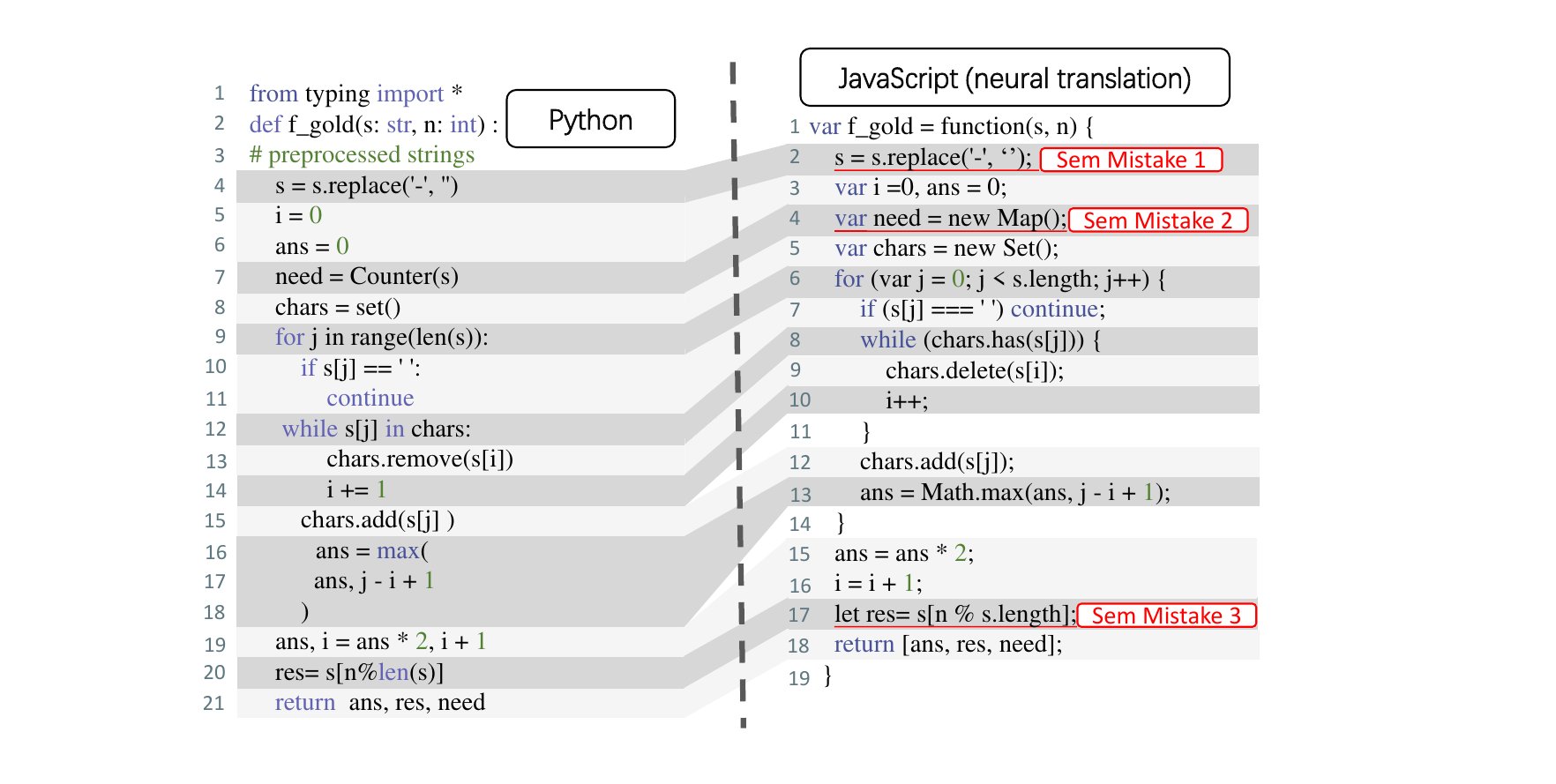}
  \caption{An example of translated code from a neural code generator. The code contains 3 semantic mistakes.}
  \label{fig:example}
\end{figure*}

\begin{figure*}[h]
  \centering
  \includegraphics[width=\linewidth]{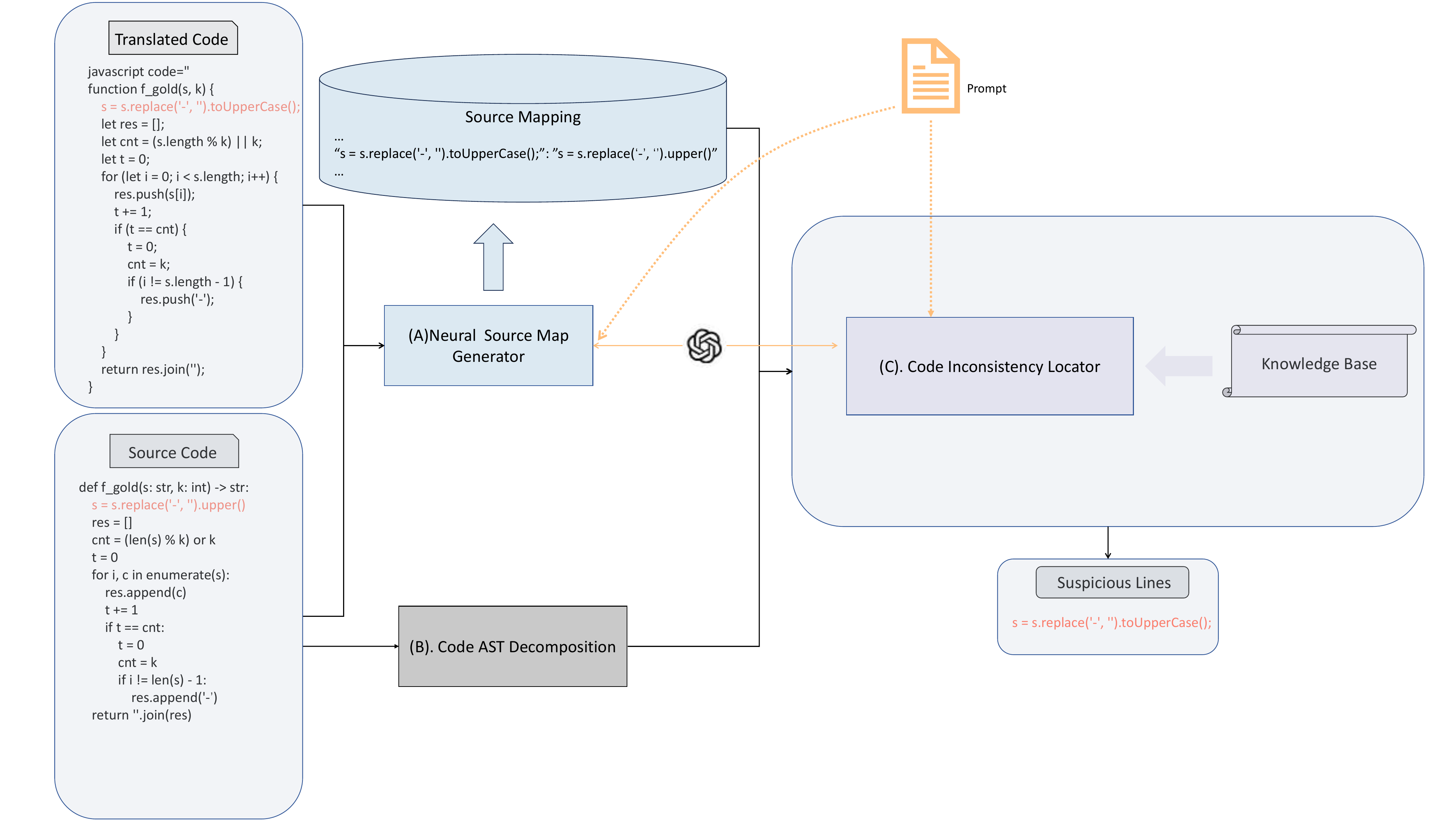}
  \caption{A static analysis framework based
on the Large Language Model (LLM). First, the framework generates a semantic mapping between source code and translated code.Next, each sub-code fragment is identified by recursively traversing the abstract syntax tree of the source code, and its corresponding translated code fragment is found through the semantic mapping. Finally, EISP connects each pair of sub-code fragments with fine-grained knowledge hints through an AI chain to assist LLMs
in discovering semantic mistakes in the translated code.}
  \label{fig:overview}
\end{figure*}
\subsection{Motivation}

In previous studies, locating semantic errors in neural code translation usually relies on executing test cases. However, recent research ~\cite{yan2023codetransocean} has shown that in the field of code translation, the approach of relying on test cases is not only costly, but may also pose a security risk. In addition, for code that is imported incompletely, even if test cases exist, the code needs to be supplemented or modified. For example, if the "Counter(s)" function is used in Python code, the import statement "from collections import Counter" must be included (see Figure \ref{fig:example}). These challenges have motivated researchers to develop a new approach to locate semantic errors in code translation through static analysis rather than test cases. In the absence of test cases, it is challenging to locate semantic errors in code translation through static analysis alone. As illustrated in Figure \ref{fig:example}. The example contains multiple errors that cannot be directly localized even when executing tests and observing runtime errors. In this case, static analysis requires not only identifying the actual execution flow of the code and possible state changes, but also understanding the semantic differences between the languages, which greatly increases the complexity and difficulty of the analysis.
Figure \ref{fig:example} shows an example of code translation, where the left side is source code derived from Python, and the right side is JavaScript code translated by Codex. Although most of the Python logic is translated correctly, when we put the two codes into a test environment and provide the same test cases, we find that the translated code still outputs different results than the source code. The reason for this is that the translated JavaScript has subtle semantic errors, which can be corrected by modifying it slightly. Figure \ref{fig:example} shows the three errors as follows:

\begin{itemize}
\item \textbf{Semantic Mistake 1}: 
The s.replace('-', '') function has different semantics and functionality in the python and JavaScript languages. In python, s.replace('-', '') will by default replace all "-" in the string s with "", whereas in JavaScript, replace by default replaces only the first matched In JavaScript, replace replaces only the first "-" that matches with "", unless you use a regular expression with the global match flag, which replaces all matched characters. So the correct translation code should be "s.replace(/-/g, '');".
\item \textbf{Semantic Mistake 2}:
In the Python code, the function of the sixth line need = Counter(s) is to count the number of occurrences of each character in the string s and save the result as a dictionary. However, there is no built-in API in JavaScript similar to Python's Counter class, so the translation to JavaScript employs a Map object for similar functionality. The original translation creates a Map object using the var need = new Map(); statement, but this method by itself does not have the capability of automatically counting the frequency of characters. The correct translation would be to count the occurrences of each character by iterating over each character in the string s and updating the Map object using loops and conditional judgment logic after the Map object is created, i.e., "var need = new Map(); for (var \_c of t) need.set(\_c, 1 + (need.has(\_c) ? need.get(\_c) : 0));".
\item \textbf{Semantic Mistake 3}: 
The function of the \% in python code is also different from the \% operator in JavaScript. In python \% always returns a non-negative remainder. The remainder always remains positive or zero, regardless of whether the divisor is positive or negative. Whereas in JavaScript the sign of the remainder returned is the same as the sign of the divisor, so if the divisor is a negative number, the remainder is negative. For example, if n is negative, in python code "n\%len(s)" is still positive, but in JavaScript code "n\% s.length" is negative. So the correct JavaScript code should be "(n\%s.length + s.length) \% s.length".
\end{itemize}

\subsection{Related Work}

{\textbf{\spaceskip=0pt Locating Semantic Errors in Code Translation.}} 
wang~\linebreak\cite{wang2023transmap} et al. are the first to work on automatically locating errors in the output of neural code translations when attempting to translate across trans-programming languages. Their inputs are source code, translation code, and their test sample code, first generating line-to-line mappings of source and translation code via LLM, then executing the source and target programs using the given test inputs, which generates execution traces, and finally its comparing these execution traces based on the mappings in order to determine where the execution state (i.e., the value of the program variable being traced) differs. However, their approach relies heavily on test samples that are capable of triggering errors and is completely inapplicable for code translation without test samples. Motivated by this observation, we propose the EISP framework, which aims to be able to statically analyze source and translated code to locate semantic errors in neural code translation in the absence of test cases.

\textbf{Code Translation.} Code translation methods fall into two main categories. The first category is based on traditional source-to-source compilers or translators (transpilers), which use program analysis techniques to convert code from one language to another, such as Java2CSharp~\cite{Java2CSharp} and Sharpen~\cite{Sharpen} to convert Java code to c\# code, and CxGo~\cite{CxGo} and C2Rust~\cite{C2Rust} convert c-language programs to Go and Rust languages, respectively. Although these types of methods enable language conversion, the code generated usually has limited readability. The second category is learning-based techniques, which include the use of lexical statistical machine translation techniques ~\cite{nguyen2013lexical,nguyen2014migrating} and tree-based neural networks ~\cite{chen2018tree} to translate Java code to C\# code. In addition, unsupervised and deep learning techniques ~\cite{lachaux2021dobf,roziere2020unsupervised} have been applied to the translation of C++, Java and Python programs. Recently, Large Language Modeling (LLM) techniques in the field of Natural Language Processing have likewise been introduced into the field of code translation, showing new possibilities and effects. For example, SteloCoder~\cite{pan2023stelocoder} has designed a decoder-only LLM based on the StarCoder framework, which is specifically designed to convert C++, C\#, JavaScript, Java, or PHP code into Python code. In addition, Other research~\cite{pan2024lost} has explored the code translation capabilities of general purpose LLMs such as GPT4~\cite{OpenAIDocumentation}, Llama 2~\cite{Llama-2}, and specialized code LLMs such as CodeGen~\cite{nijkamp2022codegen}, StarCoder~\cite{li2023starcoder}, among C, C++, Go, Java and Python. Although learning-based techniques have become mainstream approaches in the field of code translation, research ~\cite{wang2023transmap, pan2024lost} has shown that these techniques tend to suffer from erroneous noisy outputs, which may introduce semantic errors in the generated translated code that are inconsistent with the source code. Therefore, a research effort is proposed to target the localization of semantic errors in neural code translation, aiming to improve the quality of code translation and reduce error propagation.

\section{Approach}
Figure \ref{fig:overview} illustrates the general framework of our approach, which consists of an offline phase and an online phase: the construction of the knowledge base is offline, while steps A, B and C in \ref{fig:overview} are online. Specifically, we first build an API knowledge base based on the official JavaScript documentation, and then we design a framework based on knowledge-driven cue chaining and code decomposition to statically locate errors in neural network translation code.
\subsection{Knowledge Base Construction}\label{sec:Knowledge Base}
As investigated earlier, some of the errors in code translation are due to confusing the functionality of similar api's in the target language and the original language, and by providing a functional description of the api's we can improve the LLM's ability to tell the difference between similar api's, and by providing it as an external knowledge, we do not need to re-train the LLM, and can improve the LLM's ability to handle the task at a small cost by simply updating the knowledge base. Previous work ~\cite{li2018improving,ren2020demystify,ren2023misuse} demonstrated the feasibility of extracting high-quality knowledge from official documents, and we follow such an approach to mine the official JavaScript documents ~\cite{MDNJavaScript} from the API's functional description knowledge.

In this work, we collect API documents from the online resource ~\cite{MDNJavaScript} through the web crawler tool ~\cite{BeautifulSoup4}, where each API document is a crawled web page containing rich information such as the name of the API, syntax, parameters, samples, and function descriptions. We only keep semi-structured API declarations and functional descriptions, where an API declaration describes the fully qualified name of an API, which serves as a retrieval index for the knowledge base. And the functional description contains the functional logic and behavior of that API.

\subsection{Source Map Generation}\label{sec:Source Map Generation}
\begin{figure}[h]
  \centering
  \includegraphics[width=\linewidth]{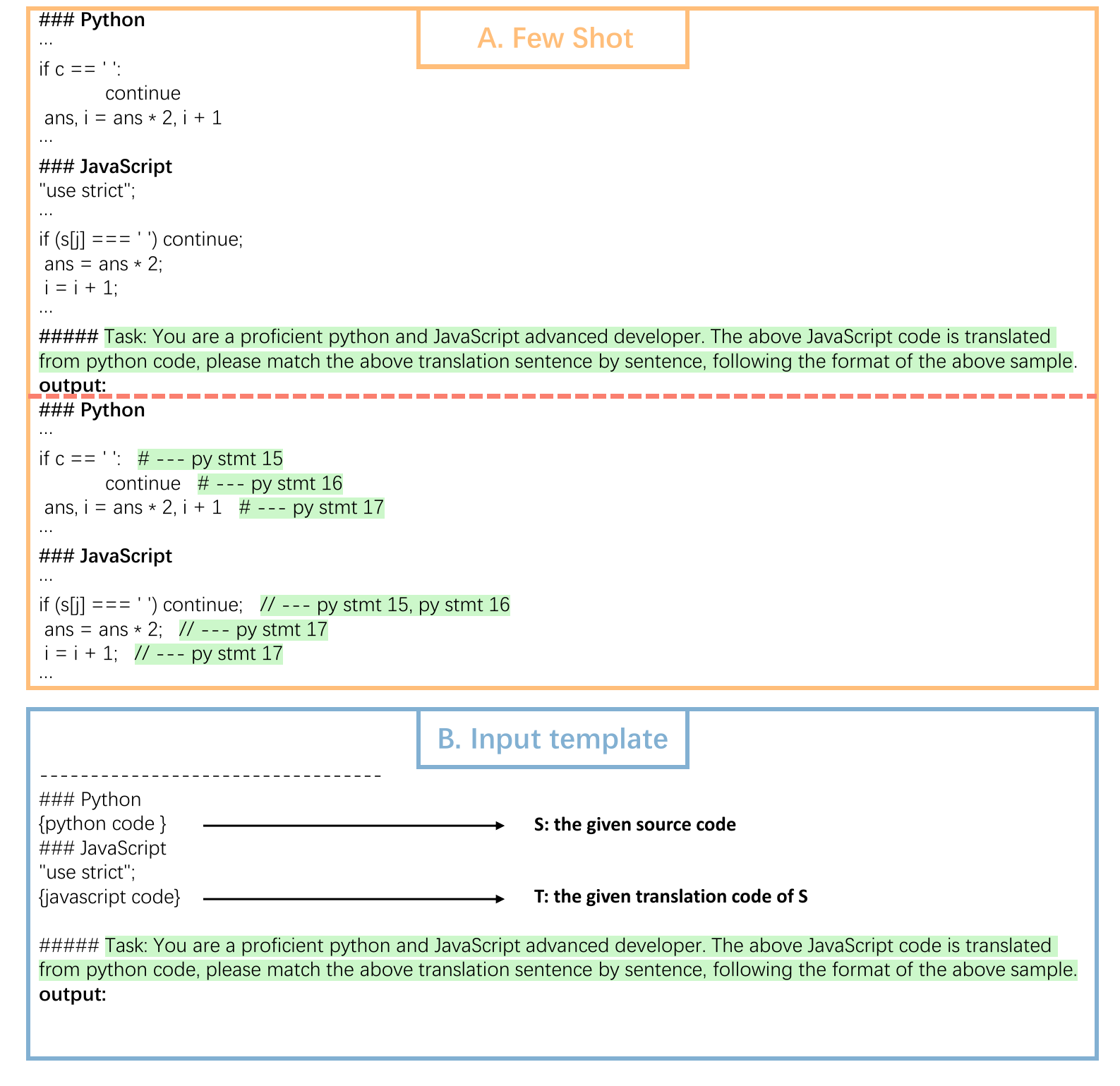}
  \caption{A prompt template for source map generation and its output.}
    \label{fig:map_prompt}

\end{figure}
In our research framework, an existing semantic mapping module is used, which aims to output the mapping between atomic fragments in the source and translated code, i.e., an ordered list of atomic fragments, based on the source and translated code (S,T) of a given neural code generator. An atomic fragment is a tuple ($(A[\ell_s], A[\ell_t])$), where $\ell_s$ and $\ell_t$ are ordered lists corresponding to line numbers in the source code and in the translation code, respectively.

This module, proposed by Wang et al.\cite{wang2023transmap}, generates source mappings by prompting LLMs using context learning. When integrating the module into our framework, inspired by the role-playing prompt proposed by ~\cite{kong2023better}, we improved the prompt of the module by prompting the LLM to play an advanced developer in python and JavaScript to improve the success rate of the mapping. The specific prompt contains two parts, the first part is the demo samples, each containing fixed python code and JavaScript code, as well as the source mapping annotations in the comments as shown in \ref{fig:map_prompt}A. Compared to the fixed first part, the second part is dynamically changing as shown in \ref{fig:map_prompt}B, where the framework replaces "{python code}" and "{ JavaScript code}" in the prompt template, and then A and B form a new prompt, prompting LLM to mimic sample A by adding comments to the code in B. The source code represents each comment in the source code. The comments in the source code represent the labeling of each line, and the comments in the translated code represent the labeling of the source code line corresponding to that line. Finally, the list of mappings is created by parsing the comments.

\subsection{AST Tree Decomposition\label{sec:AST}}
\begin{figure}[h]
  \centering
  \includegraphics[width=\linewidth]{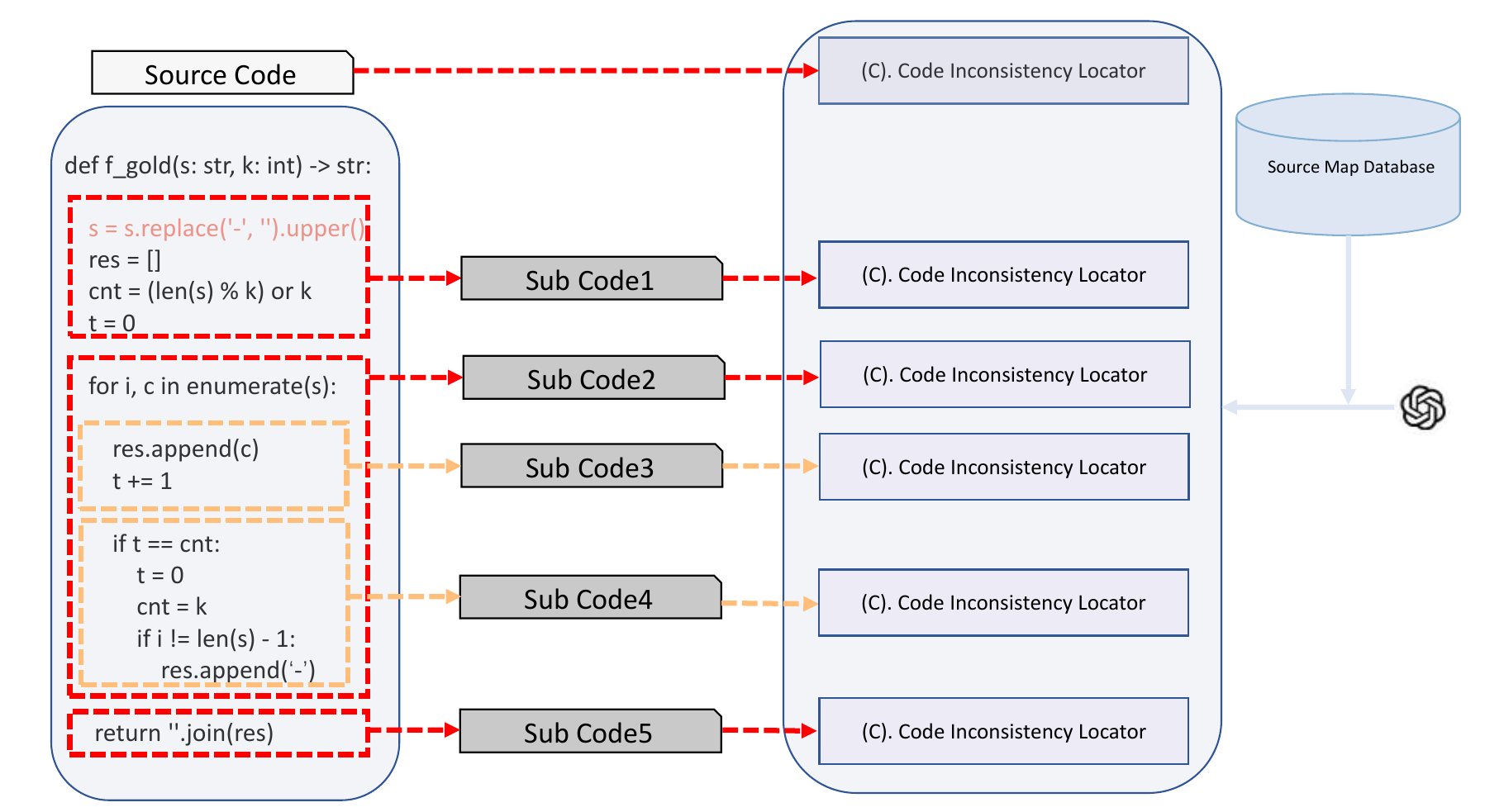}
  \caption{Demonstrates how code is recursively decomposed into sub-codes.}
    \label{fig:AST}
\end{figure}

In the process of neural code translation, we usually encounter two main problems: syntactic errors and semantic errors. For syntactic errors, large-scale language models (LLMs) can effectively identify and correct them by analyzing the overall code structure. However, for more subtle semantic errors, LLM has challenges in dealing with long texts and codes with multiple layers of nested relationships, partly due to LLM's own illusion phenomenon and distraction from long textprompts, which makes it prone to ignoring these errors.


Previous research has shown that decomposing requirements into several fine-grained sub-requirements using a divide-and-\ conquer approach can significantly improve LLM's understanding of requirements compared to dealing with coarse-grained requirements directly~\cite{ma2024compositional}. Therefore, we designed a code decomposition module as shown in Fig. \ref{fig:AST} to decompose the source code and translation code into several sub-codes synchronously and then give the sub-source code and its corresponding sub-translation code to LLM for fine-grained analysis. Regarding code decomposition, previous work ~\cite{hu2023fine,shi2023cocoast,choi2023blocsum} proposes a way to decompose code by decomposing an abstract syntax tree (AST). In AST, each "node's type name" represents a sub-fragment of the code. However, AST trees are usually large and deep ~\cite{hu2023fine,shi2023cocoast}, which are more complex than the source code, and retaining the complete AST tree is more costly in terms of space. Fine-grained decomposition on each type of node also results in a greater expenditure of time and modeling bias on the semantics of the code.

Therefore we follow the idea of Hu et al.~\cite{hu2023fine} and use eight common types of nodes in AST as predefined nodes. These eight types of nodes are: "If",
Switch", "While", "For", "Assign", "ClassDef", "Call" and "FunctionDef". By traversing the AST tree in the depth-first traversal (DFS) source code, we extract the "subtrees" under the above eight nodes as sub-code. Considering the complex nesting relationships commonly found in codes, this study continues to use the nesting depth of subcodes to measure their complexity. When the nesting depth of a subcode is below a set threshold, no further decomposition of the subcode is required. This approach ensures that the subcodes are kept at a desirable complexity level for the Large Language Model (LLM) to analyze semantic errors more accurately.

And in previous approaches, they mainly focused on decomposing a single code. In contrast, our approach requires a simultaneous decomposition of the source code and its corresponding translation code, and submits each subcode of the source code to the next processing module at the same time as the corresponding translation subcode. For this purpose, we also use the list of mappings output by the module introduced in section \ref{sec:Source Map Generation}, i.e., the source mapping database illustrated in figure \ref{fig:AST}. With this database, we are able to find the corresponding translation code for each subcode of the source code, which in turn enables the synchronized decomposition of the translation code. Thus, this method not only synchronizes the decomposition of source code and translation code, but also ensures that each sub-code can be matched one-to-one.

\subsection{Code Inconsistency Locator\label{sec:ai_chian}}       
\begin{figure*}[h]
  \centering
  \includegraphics[width=\linewidth]{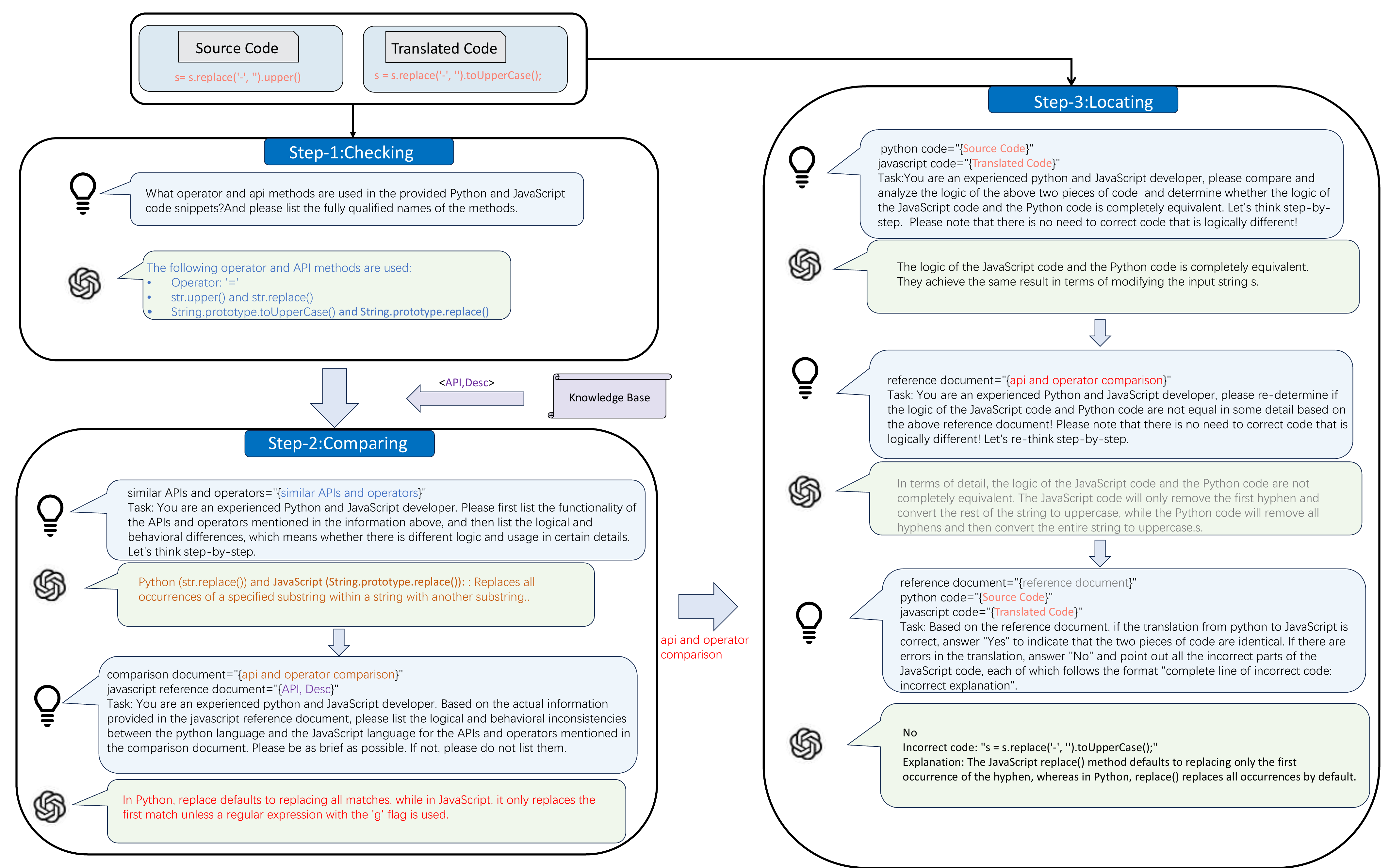}
  \caption{The AI chain flow for the Code Inconsistency Locator.}
  \label{fig:ai chain}
\end{figure*}
After obtaining the subcode and its corresponding translated code, the next step in EISP is to use LLM to statically analyze the inconsistencies between the translated code and the source code. Although LLM shows impressive potential for simple tasks, issues such as LLM's lack of transparency and insufficient controllability can degrade its performance in handling complex tasks~\cite{wu2022ai}. Therefore we have designed an AI chain workflow for this module to output the errors of inconsistency between the translated code and the source code as shown in Fig. \ref{fig:ai chain}. This AI chain consists of three steps i.e. checking, comparing and localizing which share the same LLM.

\subsubsection{detect}

The main function of the checking step is to extract fully qualified names (FQNs) of operators and API methods from source and translated code, and to query the corresponding functional descriptions through the API knowledge base. Previous research~\cite{xia2022practical, ren2023misuse} has shown that Large Language Models (LLMs) perform well in program understanding and can effectively simplify the API fully qualified name extraction process. Therefore, in order to obtain the fully qualified names of operators and APIs, we fed the source and translated code to LLM and asked with the following prompt: "What operator and api methods are used in the provided Python and JavaScript code snippets? And please list the fully qualified names of the methods."
As shown in Step 1 in Figure \ref{fig:ai chain}. Subsequently, we link the fully qualified names of these JavaScript APIs to the API knowledge base constructed in section \ref{sec:Knowledge Base} to obtain functional descriptions of the corresponding APIs.

\subsubsection{comparing}
The core purpose of the comparison step is to compare the differences in operators and APIs, such as functionality and usage, between the source code and those used in the translated code. The main challenge in this process is to utilize the accurate information in the API knowledge base to replace or correct possible misinterpretations generated by the LLM. To this end, we introduce a correction mechanism. First, we use the operators and APIs identified in Step 1 to replace the comparison phase label "{similar APIs and operators}" in Figure \ref{fig:ai chain} and prompt the LLM to perform a comparison analysis based on its in-built knowledge of the source and translated code for the corresponding operators and APIs in the source code and translated code based on its built-in knowledge. Then, the LLM is instructed to correct the results of its initial analysis by introducing the knowledge obtained from the API knowledge base, so as to accurately summarize the differences between the operators and APIs in the source and translated code.

\subsubsection{locating}

The main function of the Locate module is to identify inconsistency errors between code based on analyzing the source and translated code and the differences in operators and APIs in both. Errors in the translated code may involve not only operators and APIs, but also other types of semantic errors, such as subtle differences in looping logic~\cite{pan2024lost}. For example, the Python looping statement in the source code is: "for j in range(i - 1, 0, -1):", while the corresponding JavaScript statement generated by the neural network-driven code translation model conversion is: "for (var j = i - 1; j >= 0; j--):". This conversion leads to an important difference: the loop in the Python code stops at 1, whereas the loop in the translated JavaScript code continues to 0, introducing an inconsistent loop logic error. As shown in Step 3 in Figure \ref{fig:ai chain}, we first submit the source code and the translated code to LLM for it to initially locate all potential errors between the two. Subsequently, we provide the differences in operators and APIs summarized from the comparison step to LLM for in-depth analysis of these specific errors. Ultimately, based on the results of LLM's analysis, we will ask it whether there are any errors; if so, we will ask LLM to point out the specific lines of code that are suspicious.

\section{Experimental Design}
\subsection{Research Questions}
In the next experiment, we answer the following four main questions:


\begin{itemize}
\item \textbf{RQ1 (Effectiveness)}:  How effective is EISP at pinpointing translation mistakes?
\item \textbf{RQ2 (Vs Dynamic Analysis)}: How effective is EISP Compared to Dynamic Analysis?
\item \textbf{RQ3 (Ablation Study)}: How is the contribution of each component in pinpointing translation mistakes?
\item \textbf{RQ4 (Decomposition Thresholds)}: Does the granularity of code decomposition affect the performance of the EISP framework?
\end{itemize}

\subsection{Dataset}
As shown in Table \ref{tab:dataset}, our Python-to-Javascript benchmark contains 560 samples and 786 semantic errors. The benchmark is composed of two parts: one part is the benchmark proposed by Wang et al.~\cite{wang2023transmap}, and the other part is the extended dataset we constructed following the data generation process described by Wang et al~\cite{wang2023transmap}.
We modified the benchmark proposed by Wang et al.~\cite{wang2023transmap} by removing the test cases from both the source code and the translated code. Each sample in the benchmark contains the following elements:
\begin{itemize}
\item Source program.
\item A neural translated program with mistakes.
\item A list of mistakes in the code with line locations and fixes.
\item The fixed translated program that passes the tests.
\end{itemize}

The benchmark proposed by Wang et al. consists of 479 samples, each containing at least one error, with a total of 763 manually labeled errors—115 syntax errors and 648 semantic errors. It is based on three popular benchmarks: LeetCode Python-to-JS, GeeksForGeeks, and HumanEvalX.

To more comprehensively evaluate the effectiveness of EISP, we not only adopted the benchmark proposed by Wang et al.~\cite{wang2023transmap}, but also expanded the dataset following their benchmark construction methodology, introducing new types of semantic errors and increasing the representation of less frequent error types in the original benchmark. Specifically, using the API provided by GROQ~\cite{groqModelsDocumentation}, we employed three LLMs—gemma-7b-it, mixtral-8x7b-32768, and llama3-8b-8192—to generate 1,800 sets of translated code on the LeetCode benchmark. After manual screening and filtering, we ultimately retained 81 erroneous code samples and manually labeled 138 semantic errors within them. The criteria for manual labeling required that the translated code, if not corrected, could not pass any of the test cases on the LeetCode platform. Compared to the method used by Wang et al., which relied on a limited number of test cases to trigger errors for labeling, our approach employed significantly more comprehensive test cases. Compared to the benchmark proposed by Wang et al~\cite{wang2023transmap}. we have introduced new types of errors, such as semantic errors that cannot be triggered by the given test cases and semantic errors caused by inconsistencies in the values returned by return statements. For example, there is a logical discrepancy between Python's "return [-1, -1]" and JavaScript's "return [1, 1]", where the correct JavaScript version should be "return [-1, -1]" to maintain logical consistency with Python.


\begin{table}[ht]
\centering
\caption{Distributions of mistakes on benchmarks.}
\label{tab:dataset}
\begin{tabular}{|l|c|c|c|}
\hline
Benchmarks    & Code Count & Semantic Mistaken \\ \hline
Leetcode      & 292        & 373               \\ \hline
GeeksForGeeks & 136        & 219               \\ \hline
HumanEvalX    & 51         & 56                \\ \hline
New\_Leetcode & 81         & 138               \\ \hline
Total         & 560       & 786              \\ \hline

\end{tabular}
\end{table}

\subsection{Baselines}

Given that this study only involves source code and translation code as input data, whereas Wang et al\cite{wang2023transmap}. include both source code, translation code, and their test code in their study, a direct comparison of the two approaches may lack fairness. In addition, although both studies focus on identifying errors in neural code translation, the focus is fundamentally different. While Wang et al.'s approach relies on test cases to identify errors, this study explores an error identification approach that does not rely on test cases. Therefore, to ensure the fairness of the evaluation and to highlight the innovativeness of this study, we chose a recognized and effective prompt-based approach as the baseline for the main experiment: LLMs with Few-Shot Learning~\cite{brown2020language}: Before allowing LLMs to perform the task, a few examples are provided as demonstration examples to be added to the LLMs with Chain of Thought (CoT)~\cite{kojima2022large}: Prompt the LLM to explain the question before giving the final answer by adding "Let's think step by step" at the end of the prompt. Explain the reasoning or steps of the question before giving the final answer.

To further validate the effectiveness of our approach, we used the TransMap~\cite{wang2023transmap} method as a baseline for auxiliary experiments. This method executes both the source and target programs using given test inputs, and identifies semantic errors by comparing the execution traces, marking program statements with differing variable values as semantic errors.
\subsection{ Evaluation Metrics}
In order to evaluate the effectiveness of the EISP framework, this study follows the research metric ~\cite{wang2023transmap} of Wang et al. to examine whether EISP can identify semantic errors in benchmarks effectively. This is done by examining the suspicious lines of code identified by the EISP framework; if these lines contain errors, the detection is deemed successful. 
Here, \(\mathcal{S}_{sem}\), \(\mathcal{S}_{hid}\) and \(\mathcal{S}_{dif}\) respectively denote the proportions of successfully identified errors relative to the total semantic errors, hidden errors, and errors causing outputs different from the source code. 

\subsection{Implementation Details}

In this study, we selected GPT-3.5-turbo~\cite{OpenAIDocumentation}, GPT-4o mini~\cite{OpenAIDocumentation}, and the open-source large language model llama-3-70b-Instruct-bnb-4bit~\cite{llama3} as the base models for static analysis methods. For the dynamic analysis method, TransMap, we directly used the implementation provided by the authors. To standardize the output format, we utilized the GPT-4-1106-preview model to assist in extracting suspicious lines of code. To ensure the accuracy and stability of the results, we fixed the temperature parameter of all models to 0.2 and set the code complexity threshold to 1 to control the granularity of code decomposition.

All experiments were run on a server with 2 RTX 4090 graphics cards. To use the GPT series of models, we proceeded by directly calling the API provided by OpenAI without the need for local computational resources. This configuration not only ensures computational efficiency but also facilitates model calling and result validation through a standardized interface.
\section{Results And Analysis}
\subsection{Effectiveness of EISP in Detecting Translation Mistakes(RQ1)}

\textbf{Motivation}: 
Given that dynamic analysis methods rely on additional test cases to ensure the fairness of our experiments, this module aims to evaluate the effectiveness of EISP compared to other static analysis methods that likewise do not require test cases.

\textbf{Methodology}:
To fully leverage the potential of the base models, we selected the currently recognized prompt-based methods, CoT and Few-Shot, as baselines. Subsequently, we chose three large language models (LLMs), including two proprietary models (GPT-3.5-turbo~\cite{OpenAIDocumentation} and GPT-4o mini~\cite{OpenAIDocumentation}) and an open-source model (llama-3-70b-Instruct-bnb-4bit~\cite{llama3}), as the foundation models to validate the effectiveness and generalizability of our framework.

\textbf{Result Analysis}:
As shown in Table 2, the EISP framework based on GPT-4o mini achieved the best performance in semantic error detection, successfully identifying 82.3\% of the errors. In contrast, the CoT and Few-Shot methods using the same base model only achieved 62.0\% and 55.2\%, respectively. Moreover, when using GPT-3.5-turbo and llama-3-70b-Instruct-bnb-4bit as the base models, EISP consistently outperformed the CoT and Few-Shot methods. This indicates that our approach shows significant performance improvements compared to the baseline methods. Furthermore, EISP does not rely on a specific LLM and continues to demonstrate significant advantages across different LLMs. Additionally, EISP achieved superior performance with GPT-4o mini compared to other LLMs, possibly due to better capabilities in code analysis and instruction adherence, suggesting that the capacity of the base model has a certain impact on the effectiveness of EISP.
Moreover, we also follow the experimental setup of our previous work ~\cite{wang2023transmap} and find that the number of suspicious lines detected by our framework EISP \(\mathcal{L}_{sus}\) is 4 lines, which is on average 23\% of the number of lines of program code. This suggests that our framework allows users to understand a semantic error and then fix it by focusing on only 1 to 3 lines usually. In addition to this, we compare the performance of EISP and baselines for hidden errors\(\mathcal{S}_{hid}\) and semantic errors\(\mathcal{S}_{dif}\) that do not lead to any runtime error but make the result different from the source code. Our approach achieves an optimal success rate of 82.7\% for the former and 82.5\% for the latter. We also pay special attention to the quality of the generated source mappings; the source mappings of the previous work were 93.8\% correct, and after prompt tuning, we found that 95\% of the source mappings were correct.

\begin{table}[]
\centering
\caption{Performance of baseline methods and EISP on benchmarks.}
\label{tab:table2}
\begin{tabular}{|l|l|l|l|}
\hline
\multirow{2}{*}{Method}        & \multirow{2}{*}{\(\mathcal{S}_{sem}\)} & \multicolumn{2}{c|}{\(\mathcal{S}_{sub\_sem}\)}       \\ \cline{3-4} 
                               &                       & \multicolumn{1}{c|}{\(\mathcal{S}_{hid}\)}   & \multicolumn{1}{c|}{\(\mathcal{S}_{dif}\)}   \\ \hline
Few Shot + llama-3-70b-Instruct-bnb-4bit  & 1.0\%                & \multicolumn{1}{c|}{0.8\%} & \multicolumn{1}{c|}{1.2\%} \\ \hline
CoT + llama-3-70b-Instruct-bnb-4bit       & 8.7\%                & \multicolumn{1}{c|}{11.0\%} & \multicolumn{1}{c|}{11.1\%} \\ \hline
EISP+llama-3-70b-Instruct-bnb-4bit             & 47.1\%                & \multicolumn{1}{c|}{52.8\%} & \multicolumn{1}{c|}{56.4\%} \\ \hline

Few Shot +GPT-3.5-turbo        & 4.3\%                 & \multicolumn{1}{c|}{3.1\%}  & \multicolumn{1}{c|}{2.6\%}  \\ \hline
CoT + GPT-3.5-turbo            & 10.6\%                & \multicolumn{1}{c|}{13.0\%} & \multicolumn{1}{c|}{14.0\%} \\ \hline

EISP+GPT-3.5-turbo             & 48.7\%               & \multicolumn{1}{c|}{48.5\%} & \multicolumn{1}{c|}{48.4\%} \\ \hline

Few Shot + GPT-4O-mini             & 55.2\%                & \multicolumn{1}{c|}{56.1\%} & \multicolumn{1}{c|}{56.0\%} \\ \hline

CoT + GPT-4O-mini             & 62.0\%                & \multicolumn{1}{c|}{65.1\%} & \multicolumn{1}{c|}{64.4\%} \\ \hline

EISP+GPT-4O-mini             & \textbf{82.3\%}                & \multicolumn{1}{c|}{\textbf{82.7\%}} & \multicolumn{1}{c|}{\textbf{82.5\%}} \\ \hline

\end{tabular}
\end{table}

\begin{tcolorbox}[
    rounded corners, 
    colback=gray!20, 
    colframe=black, 
    boxrule=1pt, 
    coltext=black, 
    boxsep=5pt, 
    left=1pt, 
    right=1pt, 
    top=1pt, 
    bottom=1pt 
]

\textbf{Answer to RQ1:} EISP successfully identified 82.3\% of the semantic errors in the benchmark tests, significantly outperforming all compared baseline methods. Additionally, EISP is not dependent on a single LLM, demonstrating strong generalizability.
\end{tcolorbox}

\subsection{Effectiveness of EISP (Static Analysis) Compared to Dynamic Analysis(RQ2)}


\textbf{Motivation}: As the first approach to leverage pure static analysis for locating semantic errors in code translation, this study aims to explore whether EISP, beyond its inherent advantages of not relying on test cases and not requiring code execution, can also outperform existing dynamic analysis methods in terms of semantic error localization success rate.

\textbf{Methodology}: To address this question, we adopted the benchmark tests and evaluation metrics mentioned in the experimental setup and selected the only available dynamic analysis method, TransMap, as a baseline for comparison. Given that TransMap requires manual intervention to fix each identified error before continuing to locate subsequent errors, we also included a variant of TransMap without manual intervention (referred to as "TransMap without human") as an additional baseline. Since dynamic analysis methods depend on test cases, we provided test cases for TransMap in our experiments, while EISP did not require any test cases.For the existing data, we continued to use the test cases employed by Wang et al., while for the new data, we adopted the self-test cases from the LeetCode platform following the approach of Wang et al~\cite{wang2023transmap}.                  
\textbf{Result Analysis}: As shown in Table \ref{tab:tableRQ4}, the EISP framework achieved a score of 82.3\% on the overall semantic error metric \(\mathcal{S}_{sem}\), outperforming the baseline method TransMap by 7.74 percentage points. For the specific semantic error types \(\mathcal{S}_{hid}\) and \(\mathcal{S}_{dif}\), EISP achieved 82.7\% on 
\(\mathcal{S}_{hid}\), which is comparable to the baseline, and 82.5\% on 
\(\mathcal{S}_{dif}\), surpassing the baseline by 17.9 percentage points. Moreover, compared with the baseline method without manual intervention (TransMap without human), EISP demonstrated comprehensive superiority: outperforming by 31 percentage points on the overall semantic error metric \(\mathcal{S}_{sem}\), and by 14.8 and 30.4 percentage points on the specific semantic error metrics \(\mathcal{S}_{hid}\)and \(\mathcal{S}_{dif}\), respectively.
These experimental results indicate that EISP outperforms the baseline methods in overall semantic error localization. Particularly in handling explicit semantic errors that lead to output discrepancies, EISP demonstrated outstanding performance.  However, in handling hidden semantic errors \(\mathcal{S}_{hid}\), EISP's performance is slightly inferior to TransMap with manual intervention, but significantly better than TransMap without manual intervention. This further demonstrates that EISP, without relying on test cases or manual intervention, can still exhibit strong error localization capabilities under specific scenarios.

\begin{tcolorbox}[
    rounded corners, 
    colback=gray!20, 
    colframe=black, 
    boxrule=1pt, 
    coltext=black, 
    boxsep=5pt, 
    left=1pt, 
    right=1pt, 
    top=1pt, 
    bottom=1pt 
]




\textbf{Answer to RQ2:} EISP successfully identified 82.3\% of semantic errors in benchmark tests, significantly outperforming all existing dynamic analysis methods. Furthermore, unlike dynamic analysis methods such as Transmap, EISP does not rely on test cases or require human intervention, providing greater generalizability.
\end{tcolorbox}

\begin{table}[]
\centering
\caption{ Results of comparing EISP with the dynamic analysis method TransMap and the TransMap method without human intervention (TransMap without human).}
\label{tab:tableRQ4}
\begin{tabular}{|l|l|l|l|}
\hline
\multirow{2}{*}{Method}        & \multirow{2}{*}{\(\mathcal{S}_{sem}\)} & \multicolumn{2}{c|}{\(\mathcal{S}_{sub\_sem}\)}       \\ \cline{3-4} 
                               &                       & \multicolumn{1}{c|}{\(\mathcal{S}_{hid}\)}   & \multicolumn{1}{c|}{\(\mathcal{S}_{dif}\)}   \\ \hline
TransMap without human
 & 51.3\%                & \multicolumn{1}{c|}{67.9\%} & \multicolumn{1}{c|}{52.1\%} \\ \hline
TransMap      & 74.9\%                & \multicolumn{1}{c|}{84.4\%} & \multicolumn{1}{c|}{64.6\%} \\ \hline
EISP + GPT-4o mini             & \textbf{82.3\%}               & \multicolumn{1}{c|}{82.7\%} & \multicolumn{1}{c|}{\textbf{82.5\%}} \\ \hline
\end{tabular}
\end{table}

\subsection{Impacts of Different Modules in the EISP Framework(RQ3)}

\begin{figure}[h]
  \centering
  \includegraphics[width=\linewidth]{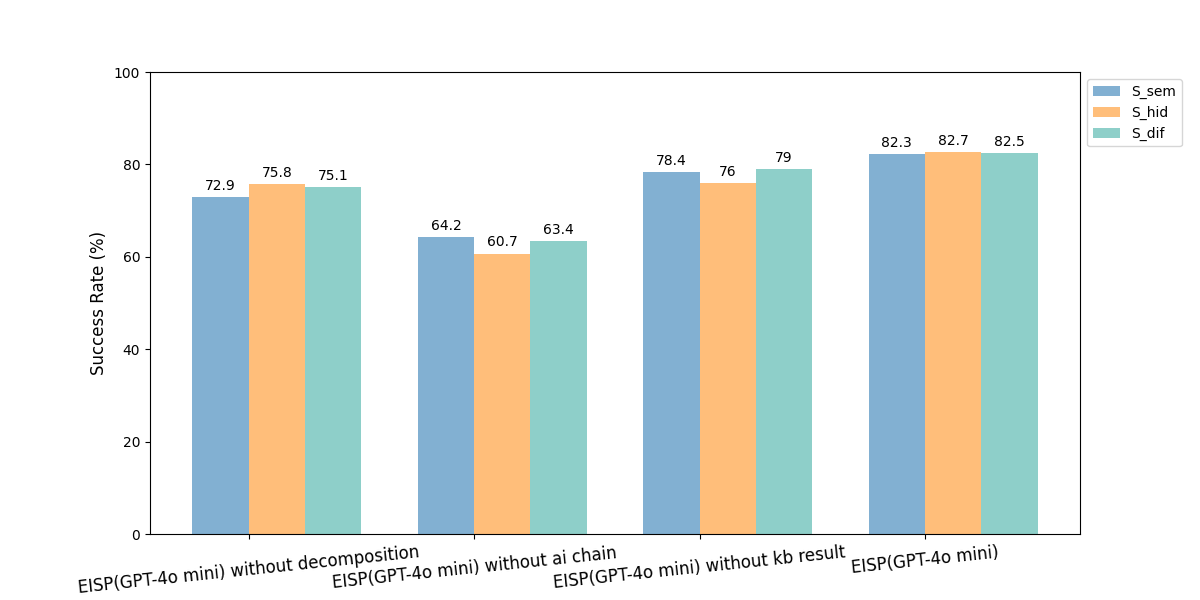}
  \caption{The figure shows the ablation results for three key components.}
    \label{fig:ablation_picture}
\end{figure}


\textbf{Motivation}: We aim to conduct ablation experiments to determine the contribution of each key component of the EISP framework.

\textbf{Methodology}: The ablation study involved three configurations: (1) EISP framework without an AST decomposition module: we removed the decomposition module and fed the source and translation code directly into the AI chain module; (2) EISP framework without an AI chain module: we replaced the AI chain component with the Few-Shot prompt; (3) EISP framework without knowledge base: we reset the knowledge queried from the API knowledge base to empty in the AI chain module to avoid impacting the AI chain module.

\textbf{Result Analysis}: The results in Figure \ref{fig:ablation_picture} show that all three components play a crucial role in the performance of EISP. In the benchmark, the addition of the AST decomposition module improves the EISP in \(\mathcal{S}_{sem}\), \(\mathcal{S}_{hid}\), \(\mathcal{S}_{dif}\) by 9.4\%, 6.9\%, and 7.4\%, respectively, a result which proves that by taking the complex code to fine-grained decomposition and then letting LLM analyze it, the performance of the model can indeed be improved. After adding the AI chain module, EISP improves 18.1\%, 22\%, and 19.1\% on \(\mathcal{S}_{sem}\), \(\mathcal{S}_{hid}\), and \(\mathcal{S}_{dif}\), respectively. 

This result shows that designing an AI chain disassembly step can improve the performance of LLM when dealing with complex tasks like locating semantic errors in neural code translation. After adding the API knowledge base module, EISP improves 3.9\%, 6.7\%, and 3.5\% on \(\mathcal{S}_{sem}\), \(\mathcal{S}_{hid}\), and \(\mathcal{S}_{dif}\), respectively. Introducing external API knowledge can help LLM improve its ability to distinguish similar API differences. The above results show that all three components have a very important impact on achieving optimal performance, with the addition of the AI chain producing the most significant improvement, which suggests that the AI chain plays the most critical role in the performance of EISP.
\begin{tcolorbox}[
    rounded corners, 
    colback=gray!20, 
    colframe=black, 
    boxrule=1pt, 
    coltext=black, 
    boxsep=5pt, 
    left=1pt, 
    right=1pt, 
    top=1pt, 
    bottom=1pt 
]

\textbf{Answer to RQ3:} In evaluating the contribution of the components of the EISP framework to its overall performance, we found that all three components are critical to the framework's performance. AI chain, in particular, plays a central role in EISP.
\end{tcolorbox}

\subsection{Impacts of Different Code Decomposition Granularity in the EISP Framework(RQ4)}

\begin{figure}[ht]
  \centering
  \includegraphics[width=\linewidth]{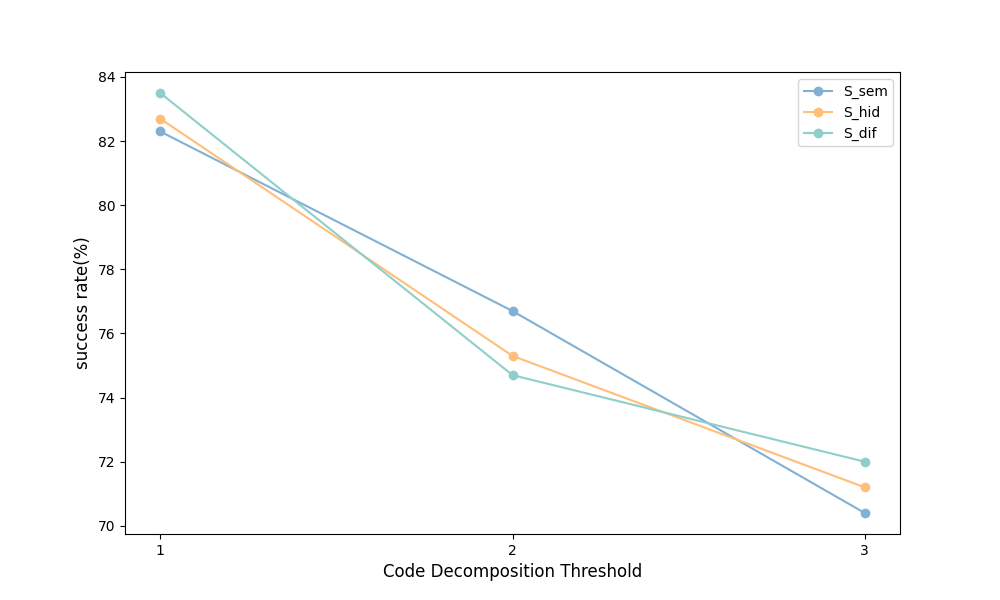}
  \caption{Effect of different decomposition thresholds on EISP performance.}
    \label{fig:ast_picture}
\end{figure}


\textbf{Motivation}:  We aim to investigate how different decomposition thresholds in the AST decomposition module impact the performance of the EISP framework, focusing on the effect of different levels of code decomposition.

\textbf{Methodology}: We evaluated the impact of varying decomposition thresholds by setting them to 1, 2, and 3. The decomposition threshold controls how deeply the code is decomposed. If the code or sub-code exceeds the specified depth, further decomposition is applied until it falls within the given threshold. This study aims to determine the optimal threshold for improving framework performance.

 \textbf{Results.} The experimental results, as shown in Figure~\ref{fig:ast_picture}, show that as the code threshold increases, the success rate in locating semantic errors~\(\mathcal{S}_{sem}\), hidden errors~\(\mathcal{S}_{hid}\), and errors that lead to a different result than the source code output~\(\mathcal{S}_{dif}\) The success rates show a general downward trend. Among them, in terms of the success rate~\(\mathcal{S}_{sem}\) in locating the semantic errors of the entire benchmark, the decomposition thresholds 1, 2, and 3 achieved 82.3\%, 76.7\%, and 70.4\%, respectively, which shows that as the code decomposition becomes more fine-grained, the simpler the code is provided to the LLM for analysis. 
 Under the influence of reducing the complex nested structure, LLM's ability to analyze the code will also increase, so the success rate of EISP in locating semantic errors in neural code translation will be higher.

\begin{tcolorbox}[
    rounded corners, 
    colback=gray!20, 
    colframe=black, 
    boxrule=1pt, 
    coltext=black, 
    boxsep=5pt, 
    left=1pt, 
    right=1pt, 
    top=1pt, 
    bottom=1pt 
]

\textbf{Answer to RQ4:} We found that the level of granularity of code decomposition is positively correlated with the ability of EISP to locate semantic errors. 

\end{tcolorbox}


\section{Threats To Validity}
In terms of validity, we have currently only investigated the localization of semantic errors in code translation from the Python language to the JavaScript language, which is also prevalent in code translation between multiple other language pairs, such as C, C++, Go, and Java with Python, according to a previous study ~\cite{pan2024lost}. Since our approach does not need to be trained for a specific language, switching to another requires only a few modifications.                        

The first step is to replace the Python and JavaScript fields in the prompt, for example, replacing Python with the source code language name and JavaScript with the translation code language name. Then, since our AST code decomposition module is based on the abstract syntax tree of the source code, switching to a different programming language only requires changing the corresponding parser, e.g., the Java parser for the Java language ~\cite{JavaParser}, the parser for the C language ~\cite{pycparser}, and so on. Finally, along the same lines, we can reconstruct the API knowledge base of the translated code from the official documentation of the corresponding language, such as the Python documentation ~\cite{PythonStdLib} and the Java documentation ~\cite{JavaSE17API}. To demonstrate the generalizability of the EISP approach, we plan to extend our research to semantic error localization in code translation for other languages in the future.

\section{Conclusion}

This paper presents an innovative approach that, for the first time, relies solely on static analysis to identify semantic errors in neural code translation without the need for test cases. 
In a series of benchmark experiments, our EISP method demonstrated significantly better performance in locating semantic errors in code translation compared to existing baseline methods. Moreover, this method showed excellent performance not only on proprietary LLM but also exhibited strong results on open-source LLMs. Additionally, compared to dynamic analysis methods, EISP offers superior performance without the need for test cases and manual intervention. Overall, this research not only provides a new solution for the software engineering community to handle code segments without test cases but also offers valuable insights into related areas such as code cloning and code searching, particularly in synchronous fine-grained decomposition and LLM application strategies.

\bibliographystyle{IEEEtran}  
\bibliography{sample-base}     


\end{document}